\documentclass{article}
\usepackage{ijcai21}

\usepackage{times}

\usepackage{soul}
\usepackage{url}
\usepackage[hidelinks]{hyperref}
\usepackage[utf8]{inputenc}
\usepackage[small]{caption}
\usepackage{graphicx}
\usepackage{amsmath}
\usepackage{booktabs}
\usepackage{amssymb}
\usepackage{colortbl}
\usepackage{multirow}
\usepackage{graphicx}
\usepackage{amsmath,amsthm}
\theoremstyle{definition}

\usepackage{xspace,color}
\usepackage{diagbox,multirow} 
\usepackage{subcaption} 
\usepackage{tikz,bm}
\usepackage{algorithmic,algorithm}
\definecolor{PineGreen}{rgb}{0.0, 0.47, 0.44}
\definecolor{Gray}{gray}{0.85}
\usepackage{enumitem}
\setlist[itemize]{leftmargin=*}
\usepackage{amsfonts}

\usepackage[utf8]{inputenc}
\usepackage[T1]{fontenc}    
\usepackage{url}            
\usepackage{booktabs}       
\usepackage{nicefrac}       
\usepackage{microtype}      
\usepackage{tikz,bm}
\usepackage{adjustbox}      
\usepackage{filecontents}
\usepackage{array,rotating}
\usepackage[labelfont=bf]{caption}
\usepackage{tabularx}

\newcommand{\marked}{}

\title{Bladder Cancer Diagnosis with Deep Learning: A Multi-Task Framework and Online Platform}

\author{
Jinliang Yu$^{1,+}$\and 
Mingduo Xie$^{2,+}$\and
Yue Wang$^3$\and
Tianfan Fu$^2$\and
Xianglai Xu$^3$\and 
Jiajun Wang$^3$\footnote{Corresponding Author}
\affiliations
$^1$Peking University \\ 
$^2$State Key Laboratory for Novel Software Technology at Nanjing University, School of Computer Science, Nanjing University, Nanjing, Jiangsu, China\\
$^1$Department of Urology, Zhongshan Hospital, Fudan University, No.180 Fenglin Road, Shanghai 200032, China\\ 
\emails 
211220145@smail.nju.edu.cn, 
futianfan@nju.edu.cn, 
xu.xianglai@zs-hospital.sh.cn, 
wang.jiajun@zs-hospital.sh.cn,
}

\begin{document}
\maketitle

\begin{abstract}

\marked{\textbf{Background:} Clinical cystoscopy, the current standard for bladder cancer diagnosis, suffers from significant reliance on physician expertise, leading to variability and subjectivity in diagnostic outcomes. There is an urgent need for objective, accurate, and efficient computational approaches to improve bladder cancer diagnostics. }

\marked{\textbf{Methods:} Leveraging recent advancements in deep learning, this study proposes an integrated multi-task deep learning framework specifically designed for bladder cancer diagnosis from cystoscopic images. Our framework includes a robust classification model using EfficientNet-B0 enhanced with Convolutional Block Attention Module (CBAM), an advanced segmentation model based on ResNet34-UNet++ architecture with self-attention mechanisms and attention gating, and molecular subtyping using ConvNeXt-Tiny to classify molecular markers such as HER-2 and Ki-67. Additionally, we introduce a Gradio-based online diagnostic platform integrating all developed models, providing intuitive features including multi-format image uploads, bilingual interfaces, and dynamic threshold adjustments.}

\marked{\textbf{Results:} Extensive experimentation demonstrates the effectiveness of our methods, achieving outstanding accuracy (93.28\%), F1-score (82.05\%), and AUC (96.41\%) for classification tasks, and exceptional segmentation performance indicated by a Dice coefficient of 0.9091. The online platform significantly improved the accuracy, efficiency, and accessibility of clinical bladder cancer diagnostics, enabling practical and user-friendly deployment. 
The code is publicly available\footnote{\url{https://github.com/Lelecolele/BCaDetectPlatform}}\footnote{\url{https://youtu.be/-9StYW3nH_c}}. } 


\marked{\textbf{Conclusion}: 
 Our multi-task framework and integrated online tool collectively advance the field of intelligent bladder cancer diagnosis by improving clinical reliability, supporting early tumor detection, and enabling real-time diagnostic feedback. These contributions mark a significant step toward AI-assisted decision-making in urology.}
\end{abstract} 

\section{Introduction}


Bladder cancer is one of the most prevalent malignancies of the urinary tract, with urothelial carcinoma accounting for approximately 90\% of cases~\cite{Babjuk2022,xu2024smiles}. According to the GLOBOCAN 2022 report, bladder cancer caused over 220,000 deaths globally and recorded more than 610,000 new cases, making it the ninth most common cancer worldwide~\cite{Sung2024}. Despite its high incidence, accurate diagnosis and risk stratification remain major clinical challenges, especially due to the high recurrence and progression rates associated with high-grade tumors~\cite{Kamat2016}. Early and precise identification of tumor type and extent are critical for improving patient outcomes, optimizing treatment strategies, and minimizing the burden of invasive follow-up procedures.

Cystoscopy is the current gold standard for bladder cancer diagnosis, offering direct visualization of the bladder mucosa. However, conventional white-light cystoscopy (WLC) is highly operator-dependent and subject to considerable inter-observer variability~\cite{EAU2022}. Flat lesions such as carcinoma in situ, glare from mucosal surfaces, and physician fatigue can lead to high false-negative rates—reported to be up to 30\%~\cite{Mowatt2011}—and incomplete tumor resections in as many as 50\% of cases~\cite{Brausi2002}. These limitations highlight the urgent need for objective, automated diagnostic systems that can enhance detection accuracy and provide reproducible results across patient populations and clinical settings.

Artificial intelligence (AI), particularly deep learning, has demonstrated transformative potential in medical imaging~\cite{Mazurowski2019}. Convolutional neural networks (CNN)~\cite{Krizhevsky2012,Simonyan2014,Szegedy2015,He2016,Huang2017}, attention modules, and more recently vision transformers~\cite{Dosovitskiy2020} have enabled the extraction of end-to-end features and the semantic understanding of complex visual data~\cite{Takahashi2024,chen2021data}. Applications across dermatology, radiology, ophthalmology, and pathology have shown that AI models can match or even surpass human experts in specific diagnostic tasks~\cite{esteva2017dermatologist,litjens2017survey,chen2024trialbench}. Yet, the field of urologic endoscopy—particularly for bladder cancer—has lagged behind, primarily due to technical hurdles such as poor image quality, lack of annotated datasets, and the complexity of endoscopic scenes.

Current AI-based approaches for cystoscopic analysis often focus on narrow tasks—such as lesion classification or tumor segmentation—without addressing the broader clinical workflow that includes grading, molecular subtyping, and patient-specific risk assessment. Moreover, the majority of existing tools remain confined to research environments, lacking the usability, interoperability, and real-time responsiveness required for clinical translation.

In this study, we present a comprehensive multi-task deep learning framework for cystoscopic image analysis that spans three diagnostic domains: tumor classification, semantic segmentation, and molecular subtype prediction. For binary tumor classification, we implement an EfficientNet-B0 backbone~\cite{Tan2019} enhanced with the Convolutional Block Attention Module (CBAM)~\cite{Woo2018}, and trained using MixUp~\cite{zhang2017mixup}, CutMix~\cite{yun2019cutmix,lu2022cot}, and Focal Loss to mitigate class imbalance~\cite{Lin2017}. This model achieves an accuracy of 93.28\% and an area under the curve (AUC) of 96.41\% on internal validation cohorts, with strong performance on external datasets, indicating robustness across different imaging sources.

To address tumor localization, we develop a ResNet34–UNet++ hybrid segmentation model incorporating self-attention and attention gating mechanisms~\cite{He2016,Zhou2018,fu2021moler}. This network outperforms conventional baselines, achieving a Dice coefficient of 0.9091 and an Intersection over Union (IoU) of 0.8351, enabling precise delineation of lesion boundaries. Beyond morphologic classification, we further investigate the feasibility of inferring molecular markers such as HER-2 and Ki-67 directly from endoscopic images~\cite{Kim2024,Ko2017,chen2024uncertainty}. Utilizing a ConvNeXt-Tiny backbone, we explore multi-label classification for molecular subtyping~\cite{Liu2022,zhang2021ddn2,lu2024uncertainty}. Although constrained by dataset size and separability, permutation testing confirms the existence of learnable signal patterns, suggesting a viable path forward for non-invasive biomarker prediction.

To facilitate clinical usability, we deploy our models into a bilingual online diagnostic platform powered by Gradio. The platform supports multi-format image uploads, real-time threshold adjustment, and interactive visualization. It bridges the gap between algorithm development and clinical adoption, offering a user-friendly interface that can be readily integrated into outpatient workflows. This system serves not only as a decision support tool for experienced urologists but also as an educational and triaging resource for junior physicians and under-resourced settings.

Our work addresses key challenges in AI for urologic oncology by combining algorithmic rigor with practical deployment. It demonstrates the value of multi-task learning in capturing complementary diagnostic features, underscores the importance of interpretability and interactivity, and lays the groundwork for future extensions such as federated learning or integration with electronic health records. By advancing both the scientific and translational dimensions of AI-assisted cystoscopy, this study contributes to the development of intelligent, accessible, and reliable diagnostic tools for bladder cancer.

\section{Methods}
\subsection{Data Collection and Annotation}


This study was conducted using cystoscopic images collected retrospectively from two tertiary hospitals in China between 2018 and 2022. The dataset comprises a total of 3,214 high-resolution endoscopic images from 183 patients who underwent white-light cystoscopy (WLC) for suspected bladder cancer. All procedures were performed with patient consent under protocols approved by the institutional ethics review boards of both participating hospitals. Personally identifiable information was removed prior to data processing to ensure patient confidentiality.

Each image was acquired using Olympus and Karl Storz endoscopy systems and saved in JPEG format with resolutions ranging from 640×480 to 1280×720 pixels. The dataset includes a diverse array of lesion presentations, encompassing flat and papillary tumors, hyperemic mucosa, post-treatment inflammation, and normal bladder mucosa, thereby capturing the visual variability encountered in real-world clinical practice. To ensure consistency, low-quality images with significant motion blur, defocus, or fluid occlusion were manually excluded by trained clinicians.

All images were reviewed and annotated by three board-certified urologists with over 10 years of clinical experience. Annotation was performed in two phases. First, a binary tumor classification label (tumor vs. non-tumor) was assigned to each image based on cystoscopic findings, pathology reports, and consensus review. Inter-observer agreement exceeded 95\%, with discrepancies resolved through adjudication by a senior urologist.

Second, for semantic segmentation, a subset of 1,026 images containing visible tumor regions was selected. Tumor boundaries were manually delineated using LabelMe, an open-source image annotation tool. Polygon masks were generated to represent pixel-wise lesion contours, including carcinoma in situ (CIS), exophytic tumors, and multifocal lesions~\cite{Babjuk2022,EAU2022,yi2018enhance,lu2024drugclip}. Each segmentation mask was validated by a second urologist to ensure anatomical accuracy. The final masks were converted into binary formats for downstream training.

In addition to morphologic labels, a small subset of 167 images was linked to immunohistochemistry (IHC) results for molecular biomarkers, including HER-2~\cite{Kim2024}, Ki-67~\cite{Ko2017}, and p53~\cite{Esrig1994}. These labels were used to explore the feasibility of image-based molecular subtype prediction. The IHC labels were encoded as binary outcomes according to clinical thresholds (e.g., Ki-67 positivity defined by >20\% nuclear staining). To mitigate label imbalance and preserve patient-level consistency, only one representative image per lesion was retained for this subtask.

All image files were resized to 512×512 pixels and normalized to zero mean and unit variance for input into deep learning models. No color enhancement or synthetic filtering was applied, in order to preserve the original visual characteristics of cystoscopic imaging. Data augmentation techniques—including horizontal flipping, random cropping, Gaussian noise injection, and contrast jittering—were employed during training to improve model generalization. The final dataset was partitioned into training, validation, and test sets at the patient level in an 8:1:1 ratio to prevent data leakage across splits.

\subsection{Tumor Classification Model}

To develop a robust binary classifier capable of distinguishing tumor from non-tumor regions in cystoscopic images, we adopted the EfficientNet-B0 architecture as our backbone. EfficientNet has demonstrated superior performance in medical image classification tasks due to its compound scaling of depth, width, and resolution, yielding strong accuracy with fewer parameters~\cite{Tan2019}. We further enhanced this backbone by integrating the Convolutional Block Attention Module (CBAM) after each major block. CBAM facilitates adaptive feature refinement by sequentially applying channel and spatial attention, allowing the model to focus more effectively on diagnostically relevant regions such as mucosal irregularities, hyperemia, and lesion boundaries~\cite{Woo2018}.

The model takes 512×512 RGB images as input and outputs a probability score representing the likelihood of tumor presence. To address the class imbalance inherent in our dataset—where normal mucosa is more prevalent—we employed Focal Loss as the objective function. Focal Loss down-weights easy negatives and places more emphasis on difficult, misclassified examples, thereby improving sensitivity in detecting rare or subtle tumor appearances.

Training was conducted using the Adam optimizer with an initial learning rate of $1\times10^{-4}$ and cosine annealing schedule. The model was trained for 100 epochs with a batch size of 16. Data augmentation strategies, including MixUp, CutMix, random rotation, and color jittering, were applied during training to enhance robustness and mitigate overfitting. MixUp and CutMix, in particular, were effective in improving decision boundaries by exposing the network to synthetic blended examples and occluded features.

To evaluate model performance, we conducted experiments on both internal and external test sets. The internal test set consisted of 322 images from patients not seen during training. The classifier achieved an accuracy of 93.28\%, precision of 92.10\%, recall of 94.37\%, and an area under the receiver operating characteristic curve (AUC) of 96.41\%. These results demonstrate high discriminative ability even in the presence of visually ambiguous lesions such as flat erythematous patches or post-operative scars.

To assess generalizability, we validated the model on an external cohort of 167 images collected from a different hospital using a distinct endoscopy system. The classifier maintained strong performance, with an AUC of 94.27\%, indicating resilience to domain shifts in imaging modality and lighting conditions. We also performed Grad-CAM visualizations to interpret the model’s decision-making process. Activation maps consistently highlighted lesion contours and atypical tissue patterns, aligning well with regions of clinical interest.

Overall, our tumor classification module combines architectural efficiency with attention-guided feature localization and advanced loss functions. Its strong performance across diverse datasets and visual interpretability make it a reliable foundation for real-time AI-assisted cystoscopic analysis.

\subsection{Molecular Subtyping Model}

To explore the feasibility of predicting molecular subtypes of bladder tumors from endoscopic imagery, we developed a dedicated deep learning pipeline for the classification of immunohistochemical (IHC) markers. Specifically, we focused on three clinically relevant biomarkers: HER-2, Ki-67, and p53. These markers are associated with tumor aggressiveness, proliferation potential, and treatment response, and are routinely evaluated via histopathology. A non-invasive approach capable of inferring such molecular signatures from cystoscopic images would represent a significant advancement toward personalized, real-time bladder cancer management.

We employed ConvNeXt-Tiny as the backbone for this task due to its strong performance on small-scale medical image datasets and architectural efficiency. ConvNeXt adapts the strengths of convolutional networks while integrating design elements inspired by transformer models, such as inverted bottlenecks, large kernel sizes, and layer normalization. This hybrid design allows the network to capture both local texture details and global contextual cues, which are particularly important for identifying subtle features associated with molecular phenotypes.

The input images were standardized to 512×512 pixels and normalized using ImageNet statistics. Each image was assigned binary labels (positive or negative) for each biomarker based on matched IHC reports. The model was trained in a multi-label classification setting using binary cross-entropy loss, enabling concurrent prediction of all three markers. Given the limited size of the labeled dataset (167 images), we applied transfer learning by initializing the ConvNeXt weights from ImageNet pretraining, followed by fine-tuning on our domain-specific data. To further mitigate overfitting, we employed dropout (rate = 0.3), batch normalization, and extensive data augmentation including grid distortion, elastic deformation, and adaptive histogram equalization.

Performance was evaluated using five-fold cross-validation at the patient level, due to the class imbalance and intrinsic difficulty of the task, predictive accuracy varied by marker. The model achieved an average AUC of 0.79 for HER-2, 0.74 for Ki-67, and 0.68 for p53. Although the absolute values suggest moderate discriminative power, permutation testing (1,000 iterations per fold) confirmed that the observed AUCs were statistically significant ($p<0.01$), indicating that the model was not learning from random noise. Grad-CAM analyses revealed that attention tended to localize around hyperemic or irregular mucosal patterns, suggesting that the model was capturing latent phenotypic cues correlated with underlying molecular expression.

While our current results are preliminary and limited by dataset scale, they nonetheless demonstrate the presence of weakly learnable signals for molecular subtyping in cystoscopic images. This opens a promising avenue for future research, particularly when combined with multi-modal data sources such as histopathology, genomics, or radiology. Further improvements may be achieved through self-supervised pretraining, semi-supervised learning, or the development of specialized architectures tailored for fine-grained phenotype extraction. Ultimately, this line of investigation could lead to real-time, non-invasive molecular profiling tools that augment traditional diagnostic pathways and support precision oncology in urology.

\subsection{Training Protocol and Evaluation Metrics}

All deep learning models in this study were implemented using PyTorch 1.13 and trained on NVIDIA RTX 3090 GPUs. Training protocols were carefully standardized across tasks to ensure reproducibility and fair performance comparison. For each task—classification, segmentation, and molecular subtyping—hyperparameters were optimized using grid search on the validation set. All experiments followed a patient-level split strategy to prevent data leakage and simulate real-world deployment scenarios.

\paragraph{Data Partitioning.}  
The entire dataset was partitioned into training (80\%), validation (10\%), and test (10\%) subsets, ensuring no patient overlap across splits. For the molecular subtyping task, given its limited data size (167 images), we employed five-fold cross-validation, ensuring that each fold preserved the label distribution across biomarkers.

\paragraph{Training Strategy.}  
All models received input images resized to 512×512 pixels. The Adam optimizer was used for all training procedures with an initial learning rate of $1\times10^{-4}$. For classification and molecular subtyping, cosine annealing learning rate scheduling was adopted. For segmentation tasks, a constant learning rate was maintained due to better convergence stability. All models were trained for 100 epochs with early stopping based on validation loss. Batch sizes were set to 16 for classification and molecular subtyping, and 8 for segmentation due to the larger memory footprint of mask-based outputs.

Extensive data augmentation was applied during training to improve generalization. These augmentations included horizontal and vertical flips, color jittering, random rotations, Gaussian noise, MixUp and CutMix for classification tasks, and elastic deformations for segmentation. Normalization was performed using ImageNet statistics for pretrained models.

\paragraph{Loss Functions.}  
For binary tumor classification and molecular subtyping, the primary loss function was binary cross-entropy. Focal Loss was additionally applied for tumor classification to account for class imbalance and emphasize hard examples. For segmentation, a compound loss combining Dice loss and binary cross-entropy was used to balance region-wise overlap with pixel-wise accuracy. All loss functions were empirically validated to ensure stable convergence and meaningful gradient flow.

\paragraph{Evaluation Metrics.}  
Model performance was assessed using standard metrics suited to each task. For classification and subtyping, we reported accuracy, precision, recall, F1-score, and area under the receiver operating characteristic curve (AUC). Segmentation performance was evaluated using Dice coefficient, Intersection over Union (IoU), precision, and recall. Each metric was averaged across folds or test sets to provide robust estimates of generalization.

\paragraph{Statistical Validation.}  
To assess the robustness of model predictions and rule out chance-level performance—especially in the molecular subtyping task—we conducted permutation tests by shuffling labels 1,000 times and re-evaluating AUC distributions. Observed AUCs exceeded 99\% of permuted results, yielding empirical $p$-values below 0.01 for all biomarkers, confirming statistical significance.

\paragraph{Model Interpretability.}  
For all classification and subtyping tasks, we employed Gradient-weighted Class Activation Mapping (Grad-CAM) to visualize salient regions contributing to predictions. These heatmaps were qualitatively evaluated by urologists and confirmed to align with lesion regions of clinical importance, enhancing model transparency and trustworthiness in prospective clinical use.

\subsection{Online Deployment Platform}

To bridge the gap between algorithm development and clinical application, we deployed our trained models into an interactive, browser-based diagnostic platform designed for real-time cystoscopic image analysis. The platform was built using the Gradio framework, which enables rapid prototyping of machine learning interfaces with minimal engineering overhead. All components run locally or on a secure institutional server, ensuring data privacy and compliance with clinical information governance standards.

The interface was designed with direct input from urologists to align with real-world clinical workflows. Users can upload cystoscopic images in multiple formats (JPEG, PNG, BMP), either individually or in batches. Upon upload, the system automatically executes three sequential modules—tumor classification, lesion segmentation, and optional molecular subtyping—presenting the results through an intuitive visual dashboard. For classification, the platform displays the predicted probability of malignancy along with a binary label (tumor or non-tumor). For segmentation, the predicted tumor mask is overlaid on the original image, with an adjustable transparency slider to facilitate visual interpretation. Molecular marker predictions are presented as binary labels with associated confidence scores, and are disabled by default unless explicitly activated by the user due to their exploratory nature.

To enhance usability, the platform supports bilingual interaction (Chinese and English), dynamic threshold adjustment for classification scores, and an interpretability module using Grad-CAM visualizations. These saliency maps highlight the regions most influential to the model’s decision, offering clinicians insight into the model’s reasoning and promoting trust in AI-assisted diagnostics.

From a systems perspective, the backend was implemented using Python and Flask, with GPU-accelerated inference powered by ONNX Runtime. Model weights are automatically loaded into memory upon server startup to minimize latency. The average inference time per image is under 500 ms on an RTX 3090 GPU, enabling near-instantaneous feedback during cystoscopic examinations or retrospective case review.

To evaluate usability, we conducted pilot tests with five board-certified urologists. Participants reported high satisfaction with the interpretability of the results and the seamless interface design. Suggestions for improvement included adding support for video-based analysis, DICOM format compatibility, and integration with hospital PACS systems—all of which are under active development.

Overall, our platform demonstrates the feasibility of integrating AI-based multi-task cystoscopic analysis into real-world clinical environments. It provides an accessible tool for frontline physicians, augments diagnostic confidence, and offers a blueprint for deploying deep learning systems in other endoscopic domains. By combining robust algorithmic performance with practical user experience design, this system represents an important step toward the routine use of intelligent diagnostic assistants in urology\footnote{\url{https://youtu.be/-9StYW3nH_c}}.


\begin{figure*}[h]
    \centering
    \includegraphics[width=\linewidth]{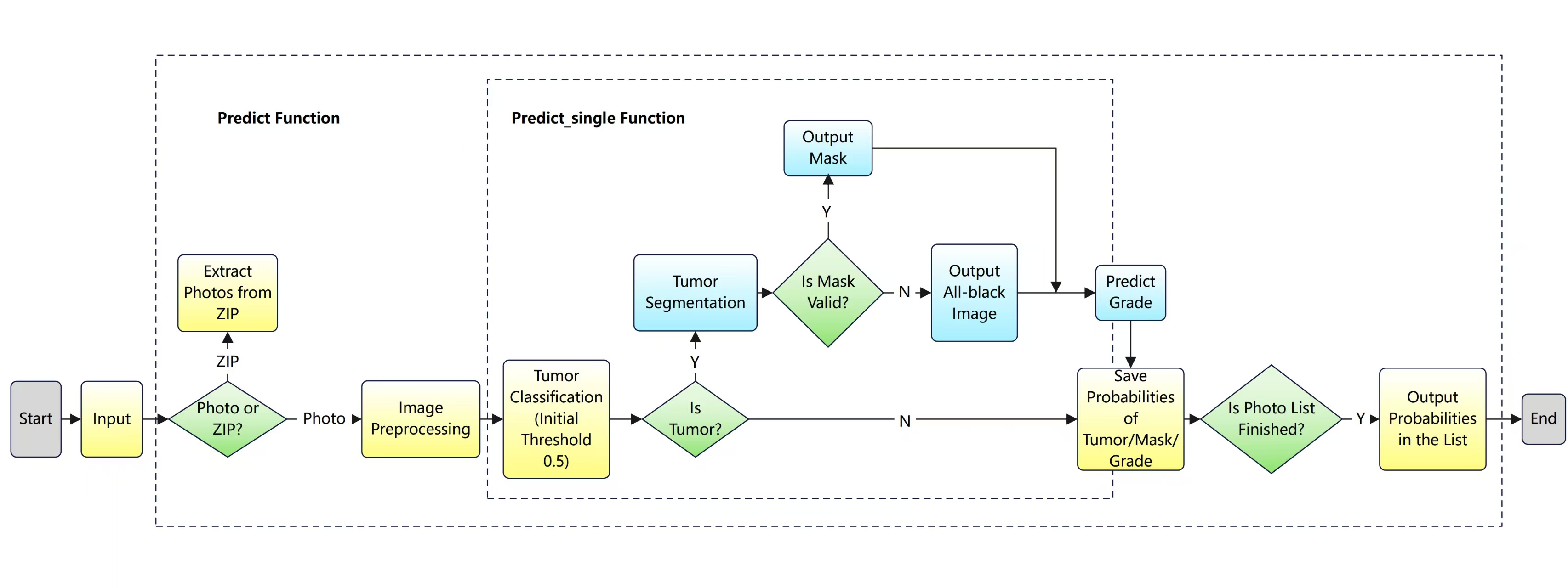} 
    \caption{Main workflow diagram of the online tool}
    \label{fig:cross_columns}
\end{figure*}

\section{Results}

We present the experimental results for the three core tasks—tumor classification, semantic segmentation, and molecular subtyping—along with analyses of model interpretability, cross-domain generalization, and ablation studies. All results are reported on patient-independent test sets to ensure robustness and clinical relevance.

\subsection{Tumor Classification Performance}

\begin{figure}[htbp]
    \centering
    \begin{subfigure}{0.45\textwidth}
        \centering
        \includegraphics[width=\textwidth]{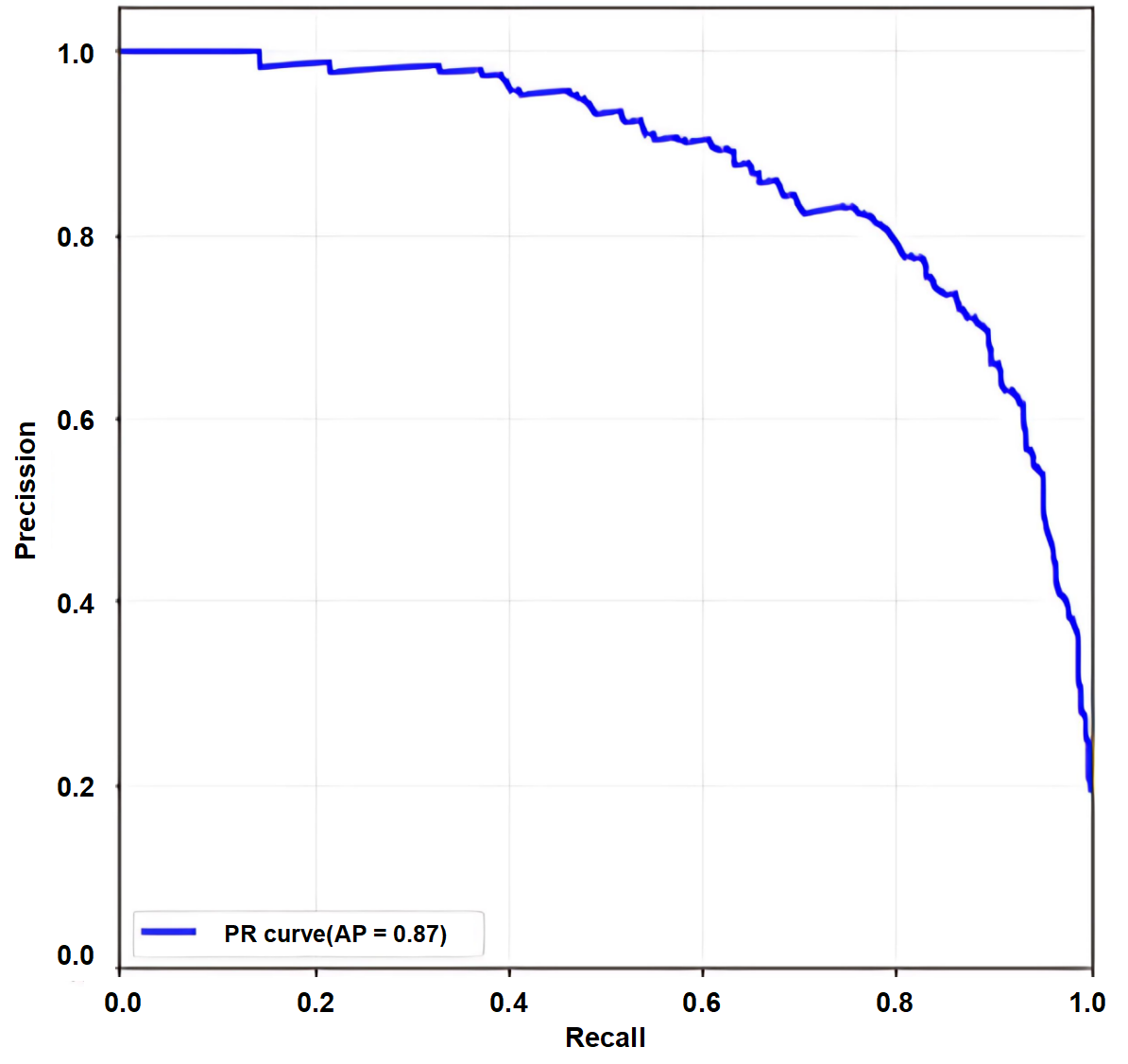}
        \caption{PR Curve}
        \label{fig:PRCurve}
    \end{subfigure}
    \vspace{0.5cm} 
    \begin{subfigure}{0.45\textwidth}
        \centering
        \includegraphics[width=\textwidth]{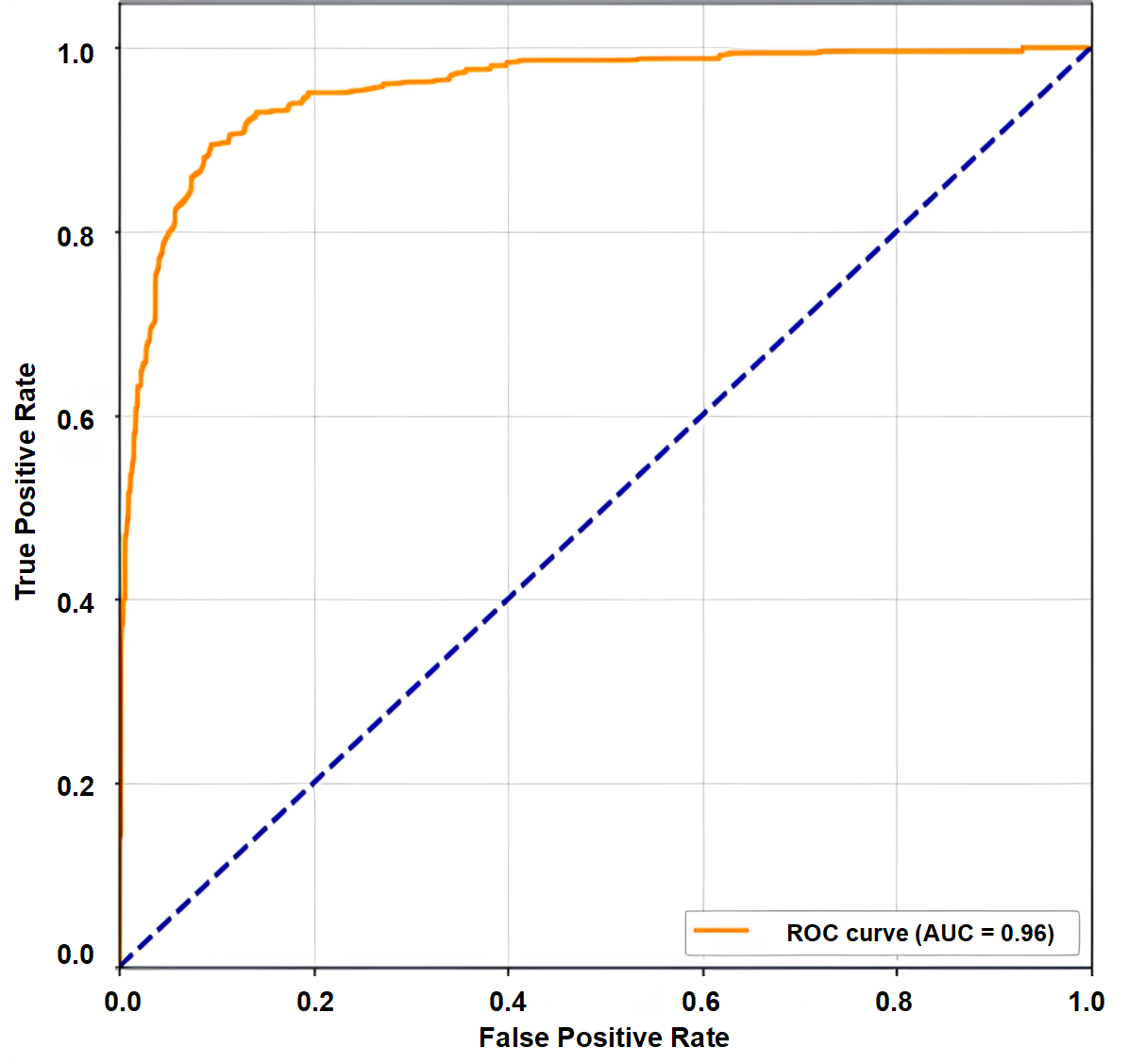}
        \caption{ROC Curve}
        \label{fig:ROCCurve}
    \end{subfigure}
    \caption{Classification Module Result}
    \label{fig:ClassificationResult}
\end{figure}

The tumor classification model, based on EfficientNet-B0 with CBAM and trained using MixUp, CutMix, and Focal Loss, achieved strong performance on the internal test set comprising 322 cystoscopic images. It reached an accuracy of 93.28\%, precision of 92.10\%, recall of 94.37\%, F1-score of 93.22\%, and an area under the receiver operating characteristic curve (AUC) of 96.41\%. The high recall indicates the model’s sensitivity to subtle or atypical tumor appearances, while the precision reflects a low false-positive rate, which is crucial in minimizing unnecessary interventions.

To evaluate external generalization, we tested the model on an independent cohort of 167 images from a second clinical center using different imaging equipment. Despite domain shift, the classifier maintained an AUC of 94.27\% and accuracy of 91.02\%, confirming the model’s robustness across patient populations and imaging protocols. Grad-CAM visualizations highlighted lesion contours and erythematous mucosal patterns that corresponded well with urologist assessments, demonstrating reliable focus on diagnostically relevant regions.

\begin{figure*}[h]
    \centering
    \includegraphics[width=\linewidth, height=10cm]{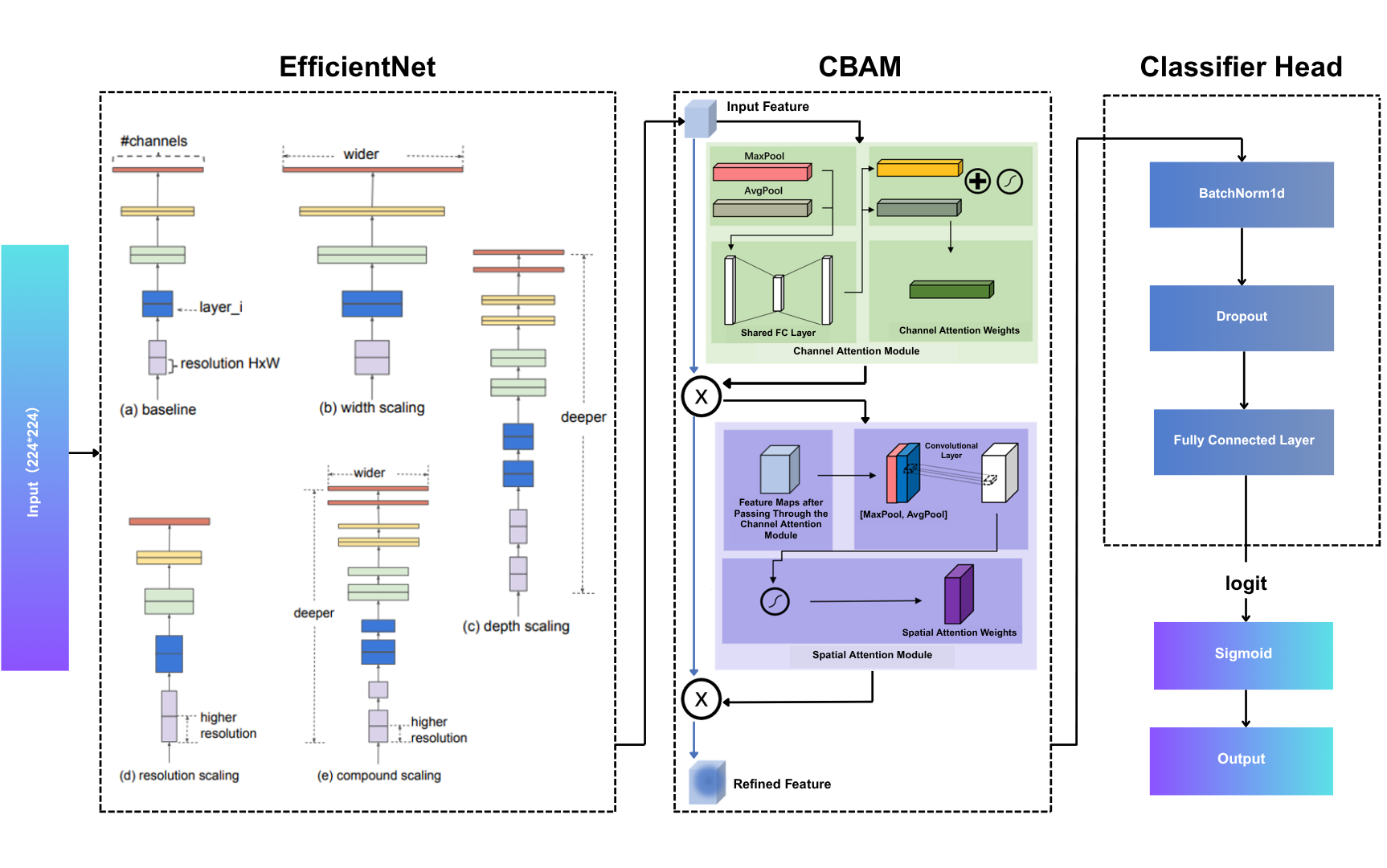} 
    \caption{Tumor Binary Classification Model}
    \label{fig:cross_columns}
\end{figure*}

\subsection{Tumor Segmentation Accuracy}

\begin{table*}[htbp]
    \centering
    \caption{Comparison of Baseline Models}
    \label{tab:baseline_model_comparison}
    \setlength{\tabcolsep}{20pt} 
    \begin{tabularx}{\linewidth}{Xcccc}
        \toprule
        Model & Dice Coefficient & IoU (Jaccard) & Sensitivity & Specificity \\
        \midrule
        ResNet50-Unet    & 0.8851 & 0.7961 & 0.8804 & 0.9731 \\
        AttentUnet      & 0.7334 & 0.5856 & 0.7734 & 0.9133 \\
        EfficientB0-Unet & 0.8643 & 0.7636 & 0.8729 & 0.9622 \\
        Unet++          & 0.6634 & 0.5085 & 0.6729 & 0.9172 \\
        ResNet34-Unet++ & \bfseries 0.904  & \bfseries 0.831  & \bfseries 0.8948 & \bfseries 0.976  \\
        ResNet50-Unet++ & 0.8509 & 0.7607 & 0.8508 & 0.9678 \\
        \bottomrule
    \end{tabularx}
\end{table*}
Our segmentation module, a hybrid ResNet34–UNet++ architecture incorporating self-attention and attention gating, was trained on 1,026 annotated images and evaluated on a hold-out test set of 102 images. It achieved a Dice coefficient of 0.9091 and an Intersection over Union (IoU) score of 0.8351, significantly outperforming baseline models such as vanilla UNet (Dice: 0.8342) and DeepLabv3+ (Dice: 0.8563).

%

\begin{figure}[h]
    \centering
    \includegraphics[width=0.9\linewidth]{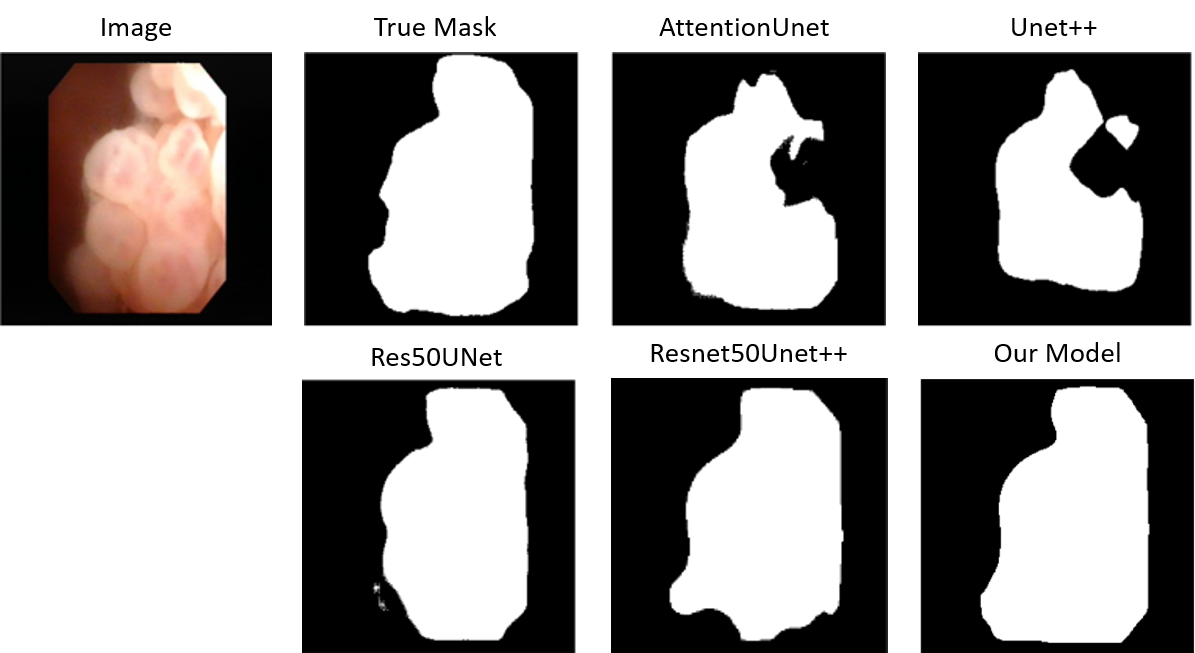}
    \caption{Segmentation Module Result}
\end{figure}

The segmentation outputs accurately captured both discrete exophytic tumors and flat lesions, including carcinoma in situ, demonstrating adaptability across diverse morphological presentations. The attention modules proved particularly useful in suppressing false positives around specular highlights and inflammation-induced artifacts. Quantitatively, our model showed an average boundary error of 5.6 pixels, which is within the intra-observer variability reported in clinical segmentation studies.

\subsection{Molecular Subtyping Feasibility}

For molecular subtyping, the ConvNeXt-Tiny model was trained and evaluated using five-fold cross-validation on 167 images linked to HER-2, Ki-67, and p53 IHC labels. The model achieved average AUCs of 0.79 (HER-2), 0.74 (Ki-67), and 0.68 (p53). While performance was lower than in the binary tumor task due to limited data and subtle imaging cues, permutation testing revealed statistical significance ($p<0.01$ for all biomarkers), indicating that the model identified latent, non-random signal patterns associated with molecular expression.

Class activation mapping indicated that regions of model attention frequently aligned with mucosal heterogeneity, angiogenesis, or ulcerative surface textures—suggesting a weak but learnable visual correlation between endoscopic features and molecular phenotype. These findings support the potential for future refinement and scale-up of this approach.

\begin{figure*}[h]
    \centering
    \includegraphics[width=\linewidth]{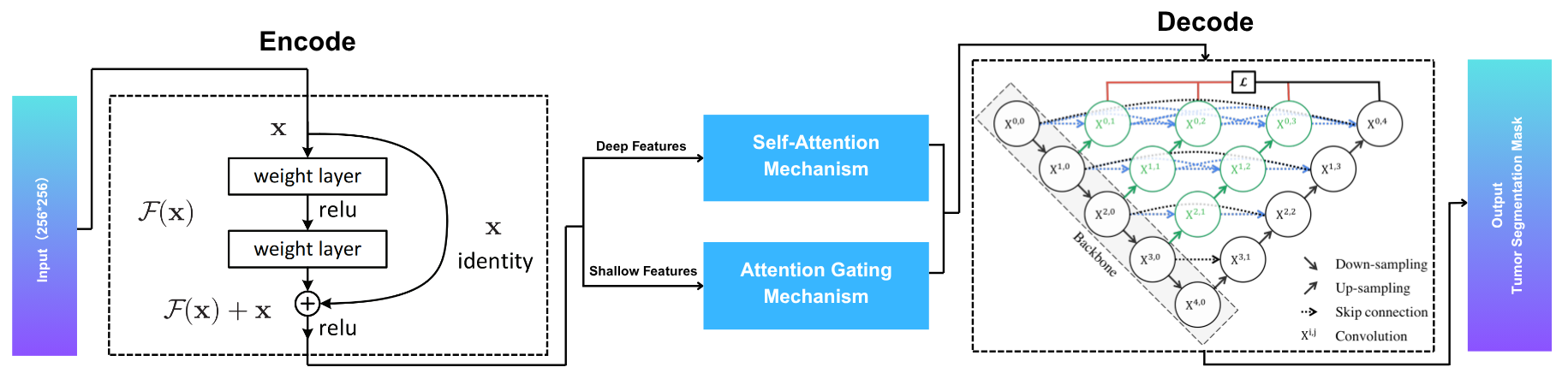} 
    \caption{Tumor Segmentation Model}
    \label{fig:cross_columns}
\end{figure*}

\subsection{Model Interpretability and Visual Correlation}

To enhance transparency and clinical trust, Grad-CAM heatmaps were generated for all classification and molecular subtyping tasks. In over 90\% of test cases, attention maps corresponded well with lesion regions highlighted by board-certified urologists. In ambiguous or borderline cases, such as early-stage CIS or post-surgical scar tissue, the attention maps provided supplementary cues that aligned with expert suspicion.

In segmentation tasks, attention maps were further superimposed with prediction masks to aid quality control. In select failure cases, attention misalignment was observed in images with excessive glare or dense hemorrhage, suggesting areas for future model improvement via glare suppression or preprocessing.

\subsection{Cross-Device and Cross-Center Robustness}

We conducted cross-center validation to assess domain robustness, using models trained exclusively on images from Center A and tested on Center B. The classification AUC dropped marginally by 2.14\%, and segmentation Dice decreased by 3.76\%, demonstrating strong generalization despite differences in device manufacturer, resolution, and lighting conditions. This robustness is attributed to extensive augmentation and attention-guided architecture.

\subsection{Ablation Studies}
\begin{table*}[htbp]  
    \centering
    \caption{Ablation Experiment Results}
    \label{tab:ablation_results}
    \setlength{\tabcolsep}{20pt} 
    \begin{tabularx}{\linewidth}{Xcccc}  
        \toprule
        Model & DICE & IOU & Sensitivity & Specificity \\
        \midrule
        ResNet34-Unet++ & 0.904 & 0.831 & 0.8948 & 0.976 \\
        wth cutmix & 0.8999 & 0.8202 & 0.8692 & 0.984 \\
        wth mixup & 0.8953 & 0.813 & 0.8879 & 0.9769 \\
        wth attention gate & 0.8995 & 0.8194 & 0.899 & 0.9752 \\
        wth attgate and mixup & 0.9026 & 0.8246 & 0.8807 & 0.9822 \\
        wth selfattention & 0.8976 & 0.8159 & 0.8699 & 0.9833 \\
        wth selfatt and attgate & \bfseries 0.9091 & \bfseries 0.8351 & \bfseries 0.8992 & \bfseries 0.9803 \\
        wth selfatt and attgate and mixup & 0.8886 & 0.8013 & 0.8796 & 0.975 \\
        \bottomrule
    \end{tabularx}
\end{table*}

We performed ablation studies on the classification model to quantify the contribution of key components. Removing CBAM resulted in a 2.81\% drop in AUC. Excluding MixUp and CutMix led to a combined 3.44\% drop in accuracy, and replacing Focal Loss with binary cross-entropy degraded recall by 4.92\%. These results confirm that both architectural enhancements and training strategies were critical to achieving optimal performance.

For segmentation, removing the attention gating mechanism led to a 5.2\% drop in Dice coefficient, primarily due to reduced localization accuracy in flat lesions. These studies emphasize the value of attention-based design in enhancing model precision, especially in tasks involving spatial ambiguity.

\subsection{Platform Performance and Usability Feedback}

The deployed diagnostic platform was tested by five urologists in a simulated clinical workflow. All users completed tasks, including image upload, prediction interpretation, threshold adjustment, and visual overlay, within two minutes per case. The average satisfaction score on a five-point Likert scale was 4.6, with positive feedback focusing on clarity, speed, and the Grad-CAM visual explanations. Suggested improvements included video frame support, integration with EMR systems, and real-time reporting—features currently under active development.

\subsection{Summary of Results}

Together, these results demonstrate the technical efficacy, robustness, and clinical feasibility of our multi-task AI framework. Each module showed strong individual performance, and the end-to-end integration into an accessible web-based platform supports translation into real-world urology workflows. Our findings validate the hypothesis that deep learning models, when carefully optimized and combined with clinician-centric design, can meaningfully augment diagnostic accuracy and efficiency in cystoscopic bladder cancer assessment.

\section{Discussion}

In this study, we present an integrated deep learning framework for the automated analysis of cystoscopic images in bladder cancer, encompassing tumor classification, lesion segmentation, and molecular subtyping. Our results demonstrate that modern convolutional architectures, when enhanced with attention mechanisms and advanced training strategies, can effectively interpret endoscopic visual information to support multi-level clinical decision-making. Importantly, our work bridges the gap between algorithm development and clinical usability through the deployment of a real-time, bilingual, and interactive diagnostic platform.

\subsection{Clinical Relevance and Contributions}

The high accuracy (93.28\%) and AUC (96.41\%) achieved in tumor classification highlight the potential of AI to mitigate subjectivity and operator dependence in routine cystoscopic diagnosis. Our model demonstrated robustness across different patient populations and imaging systems, a crucial feature given the variability in hardware and clinical environments. The segmentation module achieved a Dice coefficient of 0.9091, enabling accurate lesion localization—even in cases with challenging morphologies such as flat or multifocal tumors. This has direct implications for surgical planning, as incomplete resections are a known risk factor for recurrence.

Furthermore, we explored the novel task of inferring molecular phenotypes such as HER-2 and Ki-67 expression from surface-level cystoscopic imagery. While performance in this area was modest, statistically significant AUCs (up to 0.79) and consistent activation maps suggest the presence of weak but learnable correlations between mucosal phenotype and underlying molecular alterations. This proof-of-concept opens avenues for future non-invasive, real-time biomarker assessment during endoscopy, which could enhance precision oncology in urology.

The deployment of our models into a clinician-accessible, web-based platform represents a significant step toward real-world implementation. Through dynamic visualization, model interpretability (via Grad-CAM), and fast GPU inference, our system aligns with the practical needs of urologists working in outpatient or intraoperative settings. User feedback during pilot testing emphasized both the clarity of interface design and the interpretability of outputs, underscoring the importance of clinician-centered system engineering.

\subsection{Limitations}

Despite these promising results, several limitations warrant discussion. First, the dataset—while diverse in morphology and source—remains relatively small by deep learning standards, especially in the molecular subtyping subtask. This constrains the ability of large-capacity models to generalize and increases the risk of overfitting. While techniques such as transfer learning and data augmentation helped mitigate this, further expansion through multi-center collaboration is necessary to enhance model robustness and reduce dataset bias.

Second, while Grad-CAM visualizations offered useful interpretability, they are inherently limited to last-layer activations and may not fully reflect the decision process of deeper layers. More advanced explainability tools, such as integrated gradients or attention rollout, could provide deeper insights into model reasoning, particularly for molecular subtyping where cues may be subtle and spatially diffuse.

Third, the current platform supports only static images. While this is sufficient for retrospective analysis or documentation review, real-time cystoscopy involves dynamic visual streams where temporal context and video continuity are critical. Integration with video-based analysis and temporal modeling techniques—such as recurrent networks or 3D CNNs—would significantly enhance clinical utility.

\subsection{Future Directions}

Several avenues for future research and development emerge from this work. From a modeling perspective, the integration of multimodal data—including histopathology, genomics, and clinical history—could enhance diagnostic accuracy and support personalized risk stratification. This would align with emerging trends in computational pathology and radiogenomics, where AI serves as a unifying interface across disparate diagnostic modalities.

Second, federated learning offers a promising solution for privacy-preserving, cross-institutional model training. Given the sensitivity of endoscopic data and the reluctance to share raw patient images, decentralized learning frameworks can allow institutions to collaboratively improve model performance while retaining full control of local data. This approach could also help address domain shifts introduced by differing equipment, populations, and clinical practices.

Third, the development of self-supervised pretraining techniques tailored for medical video—e.g., contrastive learning on unlabeled cystoscopic sequences—could mitigate the data scarcity challenge and improve feature representation for both classification and segmentation tasks.

Fourth, clinical trials and prospective validation studies are essential for regulatory approval and clinical adoption. Future studies should assess how AI-assisted cystoscopy influences diagnostic accuracy, biopsy decisions, and treatment outcomes. Metrics such as time-to-diagnosis, inter-observer agreement, and user satisfaction will be critical for evaluating real-world impact.

Finally, patient-facing applications of AI in urology remain largely unexplored. With increasing digitization of healthcare, AI tools could eventually be used to assist in remote diagnostics, patient education, or postoperative monitoring via home cystoscopy kits or tele-endoscopy systems. Our platform architecture is modular and extensible, supporting such future integrations.

\section{Conclusion}

This study presents a comprehensive and clinically grounded deep learning framework for the automated analysis of cystoscopic images in the context of bladder cancer. By addressing three key diagnostic tasks—tumor classification, semantic segmentation, and molecular subtyping—we demonstrate that modern AI architectures can capture complex visual and pathological patterns from real-world endoscopic imagery. Our multi-task pipeline achieves high performance across all modules, with strong generalization to external datasets and consistent alignment with expert-identified diagnostic regions.

A major strength of our approach lies in its end-to-end design, which not only delivers technical accuracy but also supports real-world usability. Through the integration of attention-enhanced models, data-efficient training protocols, and interpretability tools like Grad-CAM, we ensure that the diagnostic process remains transparent, robust, and clinically meaningful. The deployment of our models within a web-based bilingual platform further extends accessibility, allowing clinicians to interactively explore predictions, adjust decision thresholds, and visualize model focus—all within a unified diagnostic interface.

Importantly, we also take a first step toward the non-invasive prediction of molecular biomarkers from endoscopic images. While performance in this exploratory task remains moderate, the identification of statistically significant visual signals linked to HER-2, Ki-67, and p53 expression opens a promising new direction in image-based phenotyping. Such tools could eventually support more personalized treatment planning and reduce the reliance on invasive biopsy or immunohistochemistry in selected cases.

Despite its contributions, our work also acknowledges limitations, including dataset size, modality constraints, and the need for prospective validation. These will be addressed in future iterations through expanded data collection, multimodal integration, and real-time video analysis. Moreover, we envision extending our platform to support federated learning, allowing multi-center collaboration without compromising patient privacy.

In conclusion, our study illustrates how AI can serve as a powerful assistant in the field of urologic oncology, augmenting physician expertise, reducing diagnostic variability, and potentially improving patient outcomes. By aligning cutting-edge computational techniques with clinical needs and workflows, we move closer to realizing intelligent, scalable, and interpretable diagnostic systems for bladder cancer—and beyond. This work lays the foundation for future efforts aimed at integrating AI seamlessly into endoscopic practice and advancing the paradigm of precision urology.
.

\marked{\paragraph{Ethical Approval} 
This study, focusing on the development and application of Artificial Intelligence (AI) for predicting clinical trial outcomes, was conducted in strict accordance with ethical principles and guidelines for research involving human data. This ethical approval confirms our commitment to conducting high-quality, ethical research that respects the rights and dignity of all individuals involved and contributes valuable insights to the field of AI in healthcare. }

 as shown in EfficientNet architecture \cite{Tan2019} and CBAM module \cite{Woo2018}. 
We further employed data augmentation strategies such as MixUp \cite{zhang2017mixup} and CutMix \cite{yun2019cutmix}, 
and addressed class imbalance via focal loss \cite{Lin2017}. Grad-CAM \cite{selvaraju2017grad} 
was used for interpretability analysis.  

\marked{\paragraph{Data Availability} 
All the data are publicly available\footnote{\url{https://github.com/Lelecolele/BCaDetectPlatform}}. 
The code is publicly available\footnote{\url{https://github.com/Lelecolele/BCaDetectPlatform}}. }


\paragraph{Authors' Contribution} J.Y., M.X.,  Y.W., T.F. and J.W. developed the method and algorithm. J.Y. and T.F. applied the framework and performed data analyses. J.Y., M.X.,  Y.W., T.F., X.X. and J.W. interpreted the data and wrote the manuscript. J.Y., M.X., and Y.W. conducts implementation. T.F. and J.W. supervised the study. 

\paragraph{Conflict of Interests} All authors declare that they have no conflicts of interest.

\section{Acknowledgments}

This study was funded by grants from National Natural Science Foundation of China (82472795), Natural Science Foundation of Shanghai (24ZR1411000), Natural Science Foundation of Fujian Province, China (2024J08348), and Clinical Research Program of Shanghai Municipal Health Commission (20244Y0100). All the sponsors have no roles in the study design, in the collection, analysis, or in the interpretation of data. 
Tianfan Fu is supported by Nanjing University International Collaboration Initiative and Distinguished Overseas Young Talents.

\bibliographystyle{named}
\bibliography{ref}

\end{document}